\documentclass[a4paper]{PoS}
\usepackage{bm}
\usepackage{graphicx}
\usepackage{wrapfig}

\title{Quarkonium production and TMDs at LHC}

\ShortTitle{Quarkonium and TMDs at LHC}

\author{U.~D'Alesio$^{\;a,b}$, \speaker{F.~Murgia}$^{\,\,\,\,b}$, C.~Pisano$^{\;a,b}$, and S.~Rajesh$^{\;a,b}$\\
\llap{$^a$} Dipartimento di Fisica, Universit\`a di Cagliari, Cittadella Universitaria, I-09042 Monserrato (CA), Italy\\
\llap{$^b$} INFN, Sezione di Cagliari, Cittadella Universitaria, I-09042 Monserrato (CA), Italy\\
Email:~\email{umberto.dalesio@ca.infn.it}, \email{francesco.murgia@ca.infn.it},
\email{cristian.pisano@ca.infn.it}, \email{rajesh.sangem@ca.infn.it}}

\abstract{In this contribution we briefly discuss an ongoing phenomenological programme on quarkonium
production in unpolarized and polarized proton-proton collisions in a fixed target setup at LHCb,
the LHCSpin project.
Within a TMD approach, we aim at considering in particular: the relative role of the NRQCD color-singlet
and color-octet production mechanisms, both for unpolarized and polarized quarkonium production;
the study of azimuthal and transverse single-spin asymmetries as a phenomenological tool for learning about the almost unknown gluon Sivers function; the role of initial- and final-state interactions for spin asymmetries.}

\FullConference{XXVII International Workshop on Deep-Inelastic Scattering and Related Subjects - DIS2019\\
		8-12 April, 2019\\
		Torino, Italy}

\begin{document}

\section{Introduction\label{sec:intro}}

Transverse momentum dependent distribution (TMD PDFs) and fragmentation (TMD FFs) functions
can be responsible, at the partonic level, of several spin and azimuthal asymmetries measured at hadronic level, a longstanding challenge for leading-twist, collinear perturbative QCD.
A TMD approach, based on factorization theorems, has been developed for a
class of processes with two well separated energy scales: a soft one, of the order of the intrinsic transverse motion of partons inside hadrons, and a large one required for the validity of any pQCD approach.
These processes are:
Hadron production in semi-inclusive deeply inelastic scattering (SIDIS); Drell-Yan (DY) lepton-pair production in hadronic collisions; two back-to-back hadron production in $e^+e^-$ annihilations.
Inclusive single hadron production in polarized hadronic collisions, historically
the first processes where these phenomena were observed experimentally, are the most difficult to consider. However, their experimental and theoretical study remains
crucial for, e.g., improving our knowledge on the process dependence of TMDs; investigating possible factorization breaking effects; performing a better flavor separation as compared to SIDIS and $e^+e^-$ annihilation processes.

Nowadays, RHIC at BNL is the only operating facility suitable for studying polarized spin and azimuthal asymmetries in high-energy hadronic collisions.
However, LHC at CERN naturally provides a unique opportunity for testing
TMD physics in hadronic collisions at unprecedented large center-of-mass (c.m.) energies and transverse momentum in several possible kinematical setups. Even in the absence of polarized beams or targets,
like in the present LHC configuration, many interesting azimuthal asymmetries
can be investigated in the context of TMD physics.
The provision of future LHC setups envisaging the use of polarized fixed targets could allow to extend these studies to spin asymmetries.
To this end, several proposals (AFTER@LHC, LHCSpin) have been in fact formulated in the last years.
For more detailed and updated information see e.g.~Refs.~\cite{Bjorken:2018,Aidala:2019pit} and references therein.

In the sequel we will first summarize the main physics case for the LHCSpin project, which plans to install a polarized fixed target in the LHCb experimental setup \cite{Aidala:2019pit}.
We will then briefly consider, for the $J/\psi$ case only, an ongoing project on the study of gluon TMDs through a combined analysis of gluon-dominated pion, photon, $D$ meson and quarkonium production in proton-proton collisions.

\section{The LHCSpin physics case \label{sec:phys}}

Let us now summarize the main physical motivations for the LHCSpin experimental proposal (see Ref.~\cite{Aidala:2019pit} for details). LHCSpin plans to perform:\\
 1) The study of quark TMD distributions in the almost unexplored region of medium-large quark light-cone momentum fractions, with particular interest to the Sivers function, to the transversity distribution and the tensor charge. The Boer-Mulders distribution, the Collins FF and other TMDs will be also investigated.
This physics programme requires a transversely polarized target, that will be installed in the second stage of the LHCSpin proposal. Specific processes that will be considered include, among others: 1a) Two-particle production in the same hemisphere: $pp^\uparrow\to (h_1h_2)+ X$, for the study of di-hadron FFs in a collinear factorization scheme; $pp^\uparrow \to h+{\rm jet}+ X$ and the study of azimuthal distributions of leading hadrons inside jets, TMD FFs, the Collins function and its universality properties; polarized Drell-Yan processes and the sign change of the Sivers function as compared to the SIDIS case; 1b) Two particle production in the opposite hemisphere: $pp^\uparrow\to h_1+h_2+ X$; $pp^\uparrow\to h+{\rm jet}+ X$; $pp^\uparrow\to h+\gamma+ X$.

For some of these processes TMD factorization could be violated.
Despite this, their study is fundamental for a better understanding of TMD physics. They can also be useful in order to assess the unknown relative size of factorization breaking terms in different kinematical regimes.\\
2) TMD physics in (un)polarized inclusive quarkonium production, with special emphasis on the poorly known gluon TMD distributions: the unpolarized and linearly polarized gluon TMDs in the first stage, with an unpolarized target, and the gluon Sivers function (GSF) in the proposed second stage with hydrogen and deuterium polarized targets. Specific processes of interest are:\\
2a) Quarkonium and isolated photon production in opposite hemispheres, with relative transverse momentum much smaller than the quarkonium mass $M$: $pp^\uparrow \to J/\psi + \gamma + X$, $pp^\uparrow \to \psi^\prime + \gamma + X$, $pp^\uparrow \to \Upsilon + \gamma + X$, and so on;
2b) Associated almost back-to-back quarkonium production: $pp^\uparrow \to J/\psi + J/\psi + X$,
$pp^\uparrow \to J/\psi + \psi^\prime + X$, $pp^\uparrow \to \Upsilon + \Upsilon + X$, etc.
2c) Single inclusive quarkonium, $D$ meson, pion, photon production, both in the unpolarized and in the
transversely polarized cases: $pp^\uparrow \to J/\psi, \Upsilon + X$,
$pp^\uparrow \to D(\to \mu) + X$, $pp^\uparrow \to \pi + X$, $pp^\uparrow \to \gamma + X$, and so on.

In the case of quarkonium production too, there are several open points, in particular for single inclusive particle production, which deserve further study, concerning factorization, process dependence of TMDs and their evolution with scale. Quarkonium production also requires the simultaneous use of nonrelativistic QCD (NRQCD) and the TMD approach, a subject of interest by itself. For a thorough and updated review on quarkonium production, see e.g.~Ref.~\cite{Lansberg:2019adr} and references therein.
All the LHCSpin program is clearly complementary to analogous studies undertaken at a future Electron Ion Collider (EIC), in particular concerning the universality and process dependence of TMDs.
\section{$pp^\uparrow \to J/\psi + X$: Formalism and phenomenology
\label{sec:phen}}
In this section we discuss, as an example of quarkonium physics at LHCSpin, a phenomenological study
of single inclusive $J/\psi$ production in (un)polarized  $pp$ collisions as a tool for learning
about unpolarized gluon TMDs first and the almost unknown GSF in a second step.

The theoretical approach used is the Generalized Parton Model (GPM), and, for the polarized case, also its color-gauge invariant extension (CGI-GPM), including leading-order initial- (ISIs) and final- (FSIs) state interactions. The GPM approach takes into account, assuming factorization, spin and TMD effects within the helicity formalism. As for quarkonium production, we adopt the color singlet (CS) model in a first stage. In this case, spin and azimuthal asymmetries will be independent of the unique soft long-distance matrix element (LDME) involved.

We first consider the $J/\psi$ low transverse momentum ($p_T$) spectrum, to gain information on the TMD unpolarized gluon distributions. Notice that our goal is to find reasonable agreement with available experimental results. A real fit to data would require a more sophisticated and complete next-to-leading order (NLO) analysis and large logarithms resummation.
We are mainly interested in transverse single-spin  (SSAs) and azimuthal asymmetries, where many theoretical uncertainties, of both the TMD and the NRQCD approaches, partially cancel out in the ratios of cross sections. Specifically, we aim at constraining the almost unknown gluon Sivers function and, at a later stage, study the role of intrinsic transverse motion and TMD physics in the $J/\psi$ polarization.

The process under study is
$p(p_A)\,{+}\,p(p_B)\,\to\, {\cal Q} (p_{\cal Q}) \, {+}  \,X$,
that, when the gluon fusion mechanism dominates, at partonic level corresponds to
$g(p_a)\,{+}\,g(p_b)\,\to\, Q \overline Q  [^3S^{(1)}_1] (p_{\cal Q})\,{+}\,g(p_g)\,$.

The unpolarized cross section
reads schematically (see Ref.~\cite{DAlesio:2017rzj} for full details):
\begin{equation}\label{eq:unp}
{\rm d}\sigma \, \propto \,
f_{g/p}(x_a,  k_{\perp a})
\otimes f_{g/p}(x_b, k_{\perp b}) \otimes H_{gg\to J/\psi g}^U(\hat s, \hat t, \hat u)\,,
\end{equation}
where the symbols $\otimes$ stand for a convolution over transverse momenta and light-cone momentum fractions, and the hard scattering cross section is (including the $J/\psi$ LDME $|R_0(0)|$):
\begin{equation}\label{eq:HU}
H^U_{gg\to J/ \psi g}  = \frac{5}{9}\, \vert R_0(0)\vert^2 \, M\,
\frac{\hat s^2 (\hat s-M^2)^2 + \hat t^2 (\hat t-M^2)^2 + \hat u^2 (\hat u -M^2)^2 }
{(\hat s -M^2)^2 (\hat t -M^2)^2 (\hat u -M^2)^2}\,.
\end{equation}
In the GPM, the numerator of the SSA, $({\rm d}\sigma^\uparrow - {\rm d}\sigma^\downarrow)/({\rm d}\sigma^\uparrow + {\rm d}\sigma^\downarrow) \equiv {\rm d}\Delta\sigma/(2{\rm d}\sigma)$, reads
\begin{equation}
 {\rm d}\Delta\sigma^{\rm GPM} \> \propto \>
 \left( - \frac{k_{\perp\,a}}{M_p} \right) f_{1T}^{\perp\,g}(x_a,  k_{\perp a}) \cos\phi_a
\otimes
f_{g/p}(x_b, k_{\perp b}) \otimes H_{gg\to J/\psi g}^U(\hat s, \hat t, \hat u) \,,
\label{eq:dDs-gpm}
\end{equation}
where $M_p$ is the proton mass, while in the color-gauge invariant GPM it can be written as
\begin{equation}
 {\rm d}\Delta\sigma^{\rm CGI} \> \propto \>
\left  (- \frac{k_{\perp\,a}}{M_p}\right )
 f_{1T}^{\perp\,g\,(f)}(x_a,  k_{\perp a})\cos\phi_a
\otimes f_{g/p}(x_b, k_{\perp b}) \otimes
 \left(-\frac{1}{2}\, H_{gg\to J/\psi g}^{U}(\hat s, \hat t, \hat u)\,\right)\,,
  \label{eq:dDs-cgi}
\end{equation}
where $f_{1T}^{\perp\,g\,(f)}$ is the $f$-type GSF (totally color antisymmetric, even under $C$-conjugation). The other independent GSF, the $d$-type one (totally color symmetric, odd under $C$-conjugation) does not play a role in the CS model; this case is therefore useful in disentangling the two GSFs. Notice the $(-1/2)$ factor appearing in the hard cross section, as compared to the GPM case.

\begin{wrapfigure}{l}{6.0truecm}
\vspace{-8truept}
\includegraphics[width=6.0truecm,angle=0]{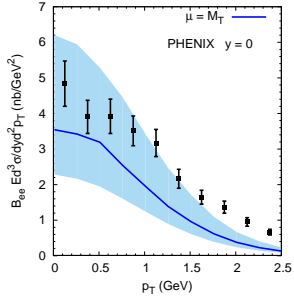}
\vspace{-8truept}
\caption{\small Unpolarized cross section for $p p\to J/\psi + X\,$ at $\sqrt{s}=200$ GeV, vs.~$p_T$ at
rapidity $y=0$. Data are from \cite{Adare:2009js}.}
\label{fig:Jcross}
\end{wrapfigure}

We adopt a flavour-independent, Gaussian shape for the unpolarized gluon distribution,
\begin{equation}\label{eq:gsfunp}
f_{g/p}(x, k_\perp,\mu) = f_{g/p}(x,\mu) \, \frac{1}{\pi \langle k_\perp^2 \rangle} \,
e^{-k_\perp ^2/\langle k_\perp^2 \rangle}\, , \qquad {\rm with} \qquad \langle k_\perp^2 \rangle = 1\,\;{\rm GeV}^2\,,
\end{equation}
taking $M_T/2 \leq \mu \leq 2M_T$ as hard factorization scale, with  $M_T = \sqrt{\bm{p}_T^2+M^2}\,$.
For the $J/\psi$, we take $M=3.097$ GeV, $|R_0(0)|^2 = 1.01$ GeV$^3$, ${\rm Br}(J/\psi \to e^+e^-) = 0.0597$.
For the $f$-type GSF, we use the positivity bound for its collinear part and fix its $k_\perp$-dependence as to maximize the effect~\cite{DAlesio:2017rzj}.
In Fig.~\ref{fig:Jcross} we compare our results for the unpolarized cross section with experimental data by the PHENIX Collaboration at RHIC~\cite{Adare:2009js} for $J/\psi$ production at central rapidity, as a function of the $J/\psi$ transverse momentum. Data include feed-down contributions; the expected fraction of direct production is 0.58.

\begin{figure}[]
\begin{center}
\includegraphics[width=5.0truecm,angle=0]{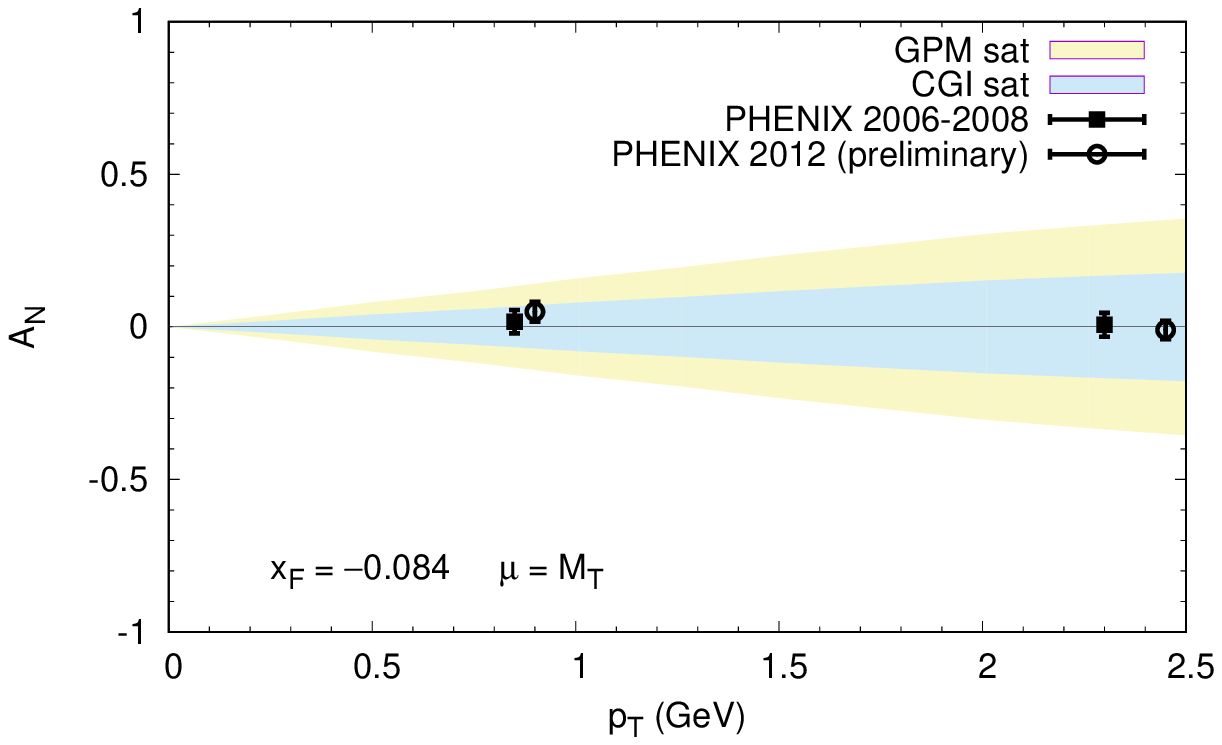}
\includegraphics[width=5.0truecm,angle=0]{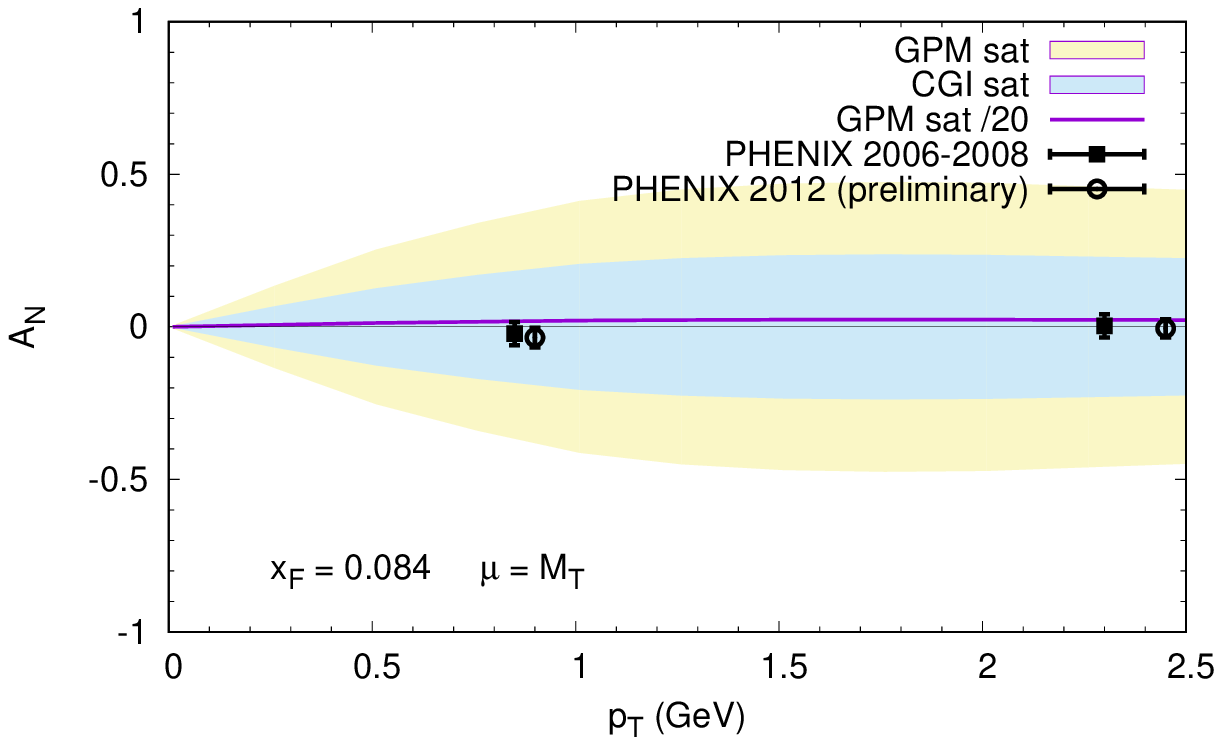}
\\
\vspace*{-16truept}
\includegraphics[width=5.0truecm,angle=0]{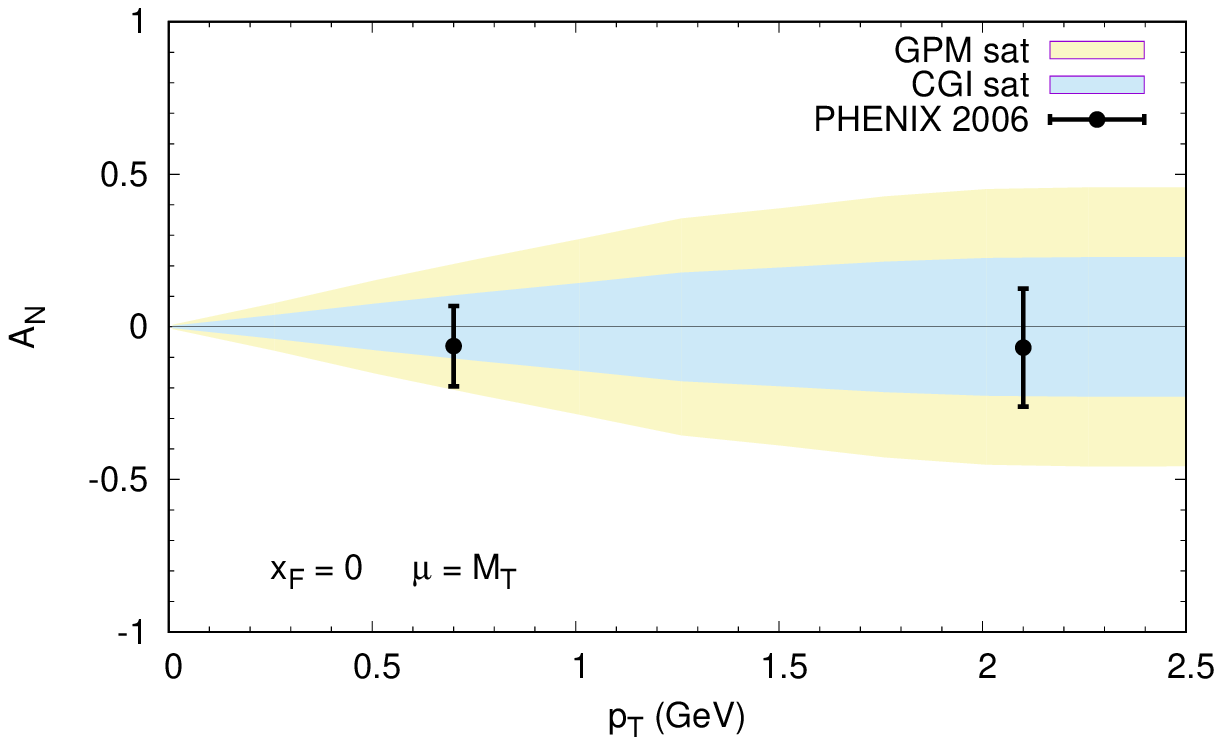}
\includegraphics[width=5.0truecm,angle=0]{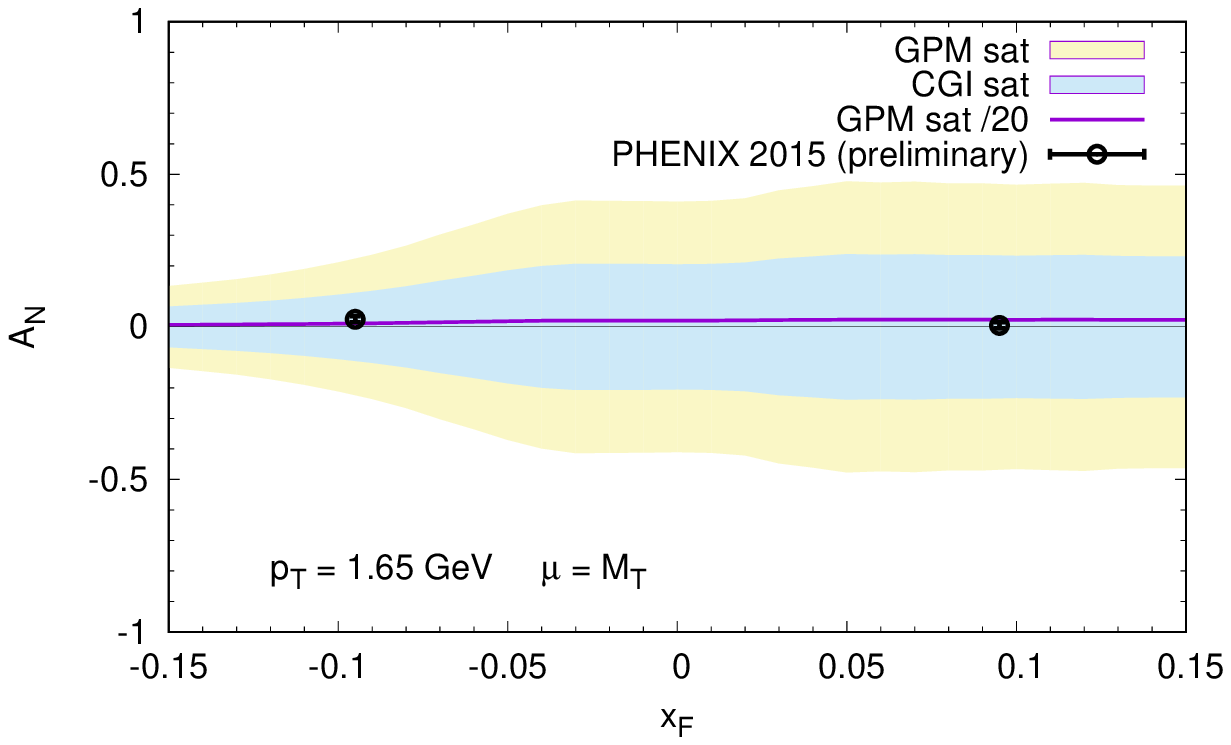}\\
\vspace*{0truept}
\caption{Bands of allowed values for $p^\uparrow p\to J/\psi+X$ Sivers SSA at $\sqrt{s}=200$ GeV in the GPM and CGI-GPM schemes, compared to PHENIX data \cite{Adare:2010bd}. See text for details.}
\label{fig:max}
\end{center}
\end{figure}

\begin{figure}[]
\begin{center}
\includegraphics[width=6.8truecm,angle=0]{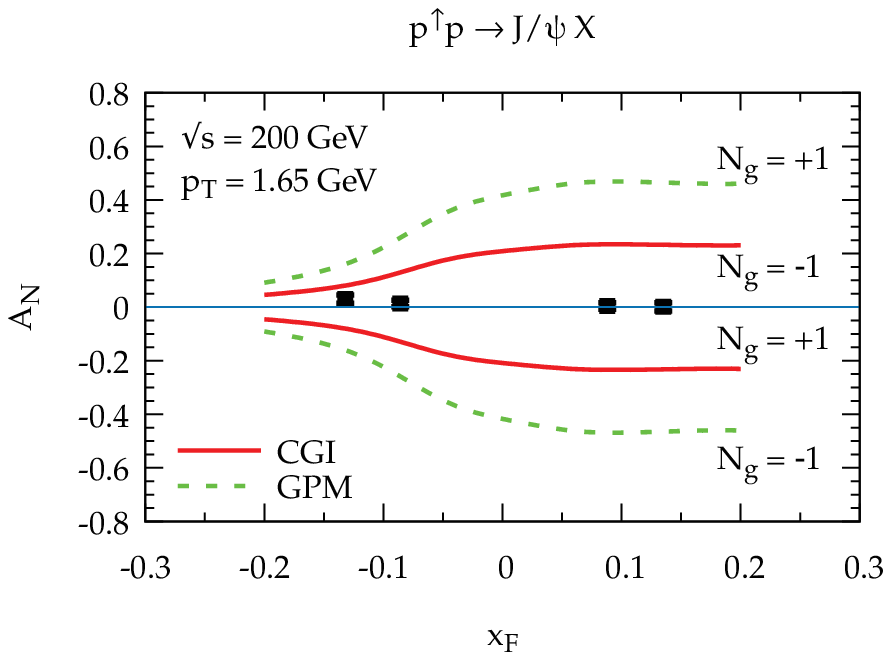}
\hspace{-60pt}
\includegraphics[width=6.8truecm,angle=0]{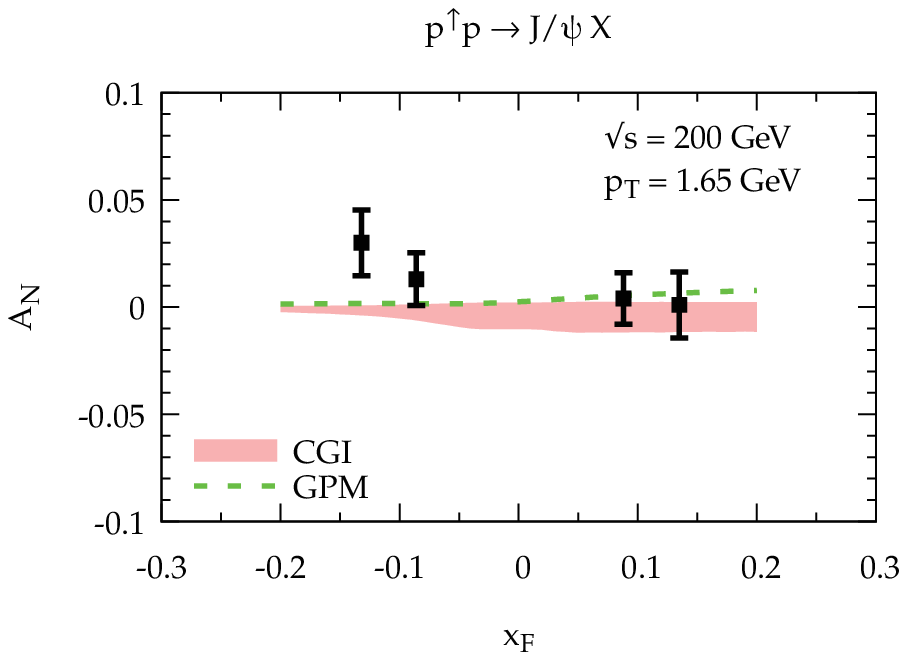}
\\
\vspace*{-10truept}
\includegraphics[width=6.8truecm,angle=0]{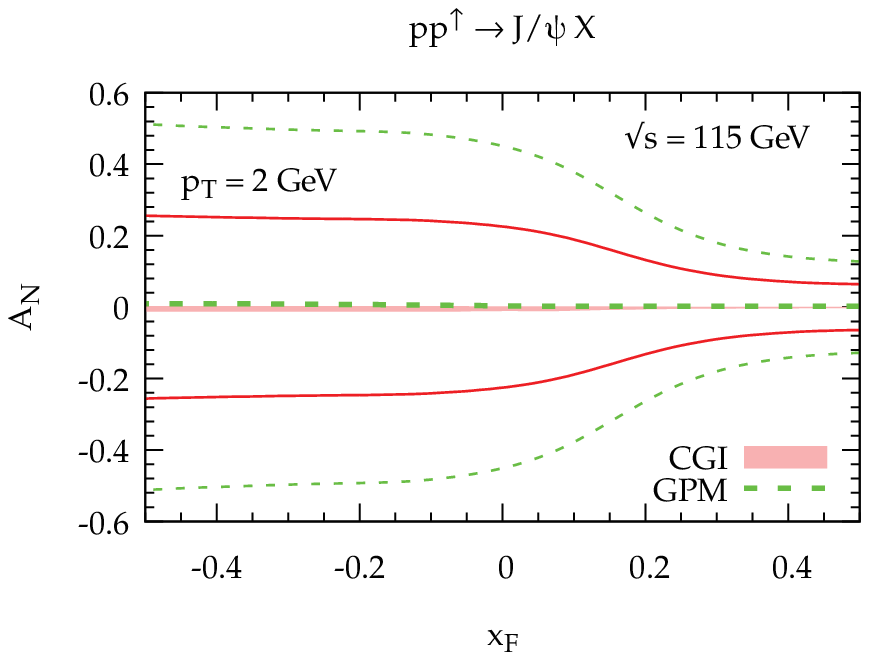}
\hspace{-60truept}
\includegraphics[width=6.8truecm,angle=0]{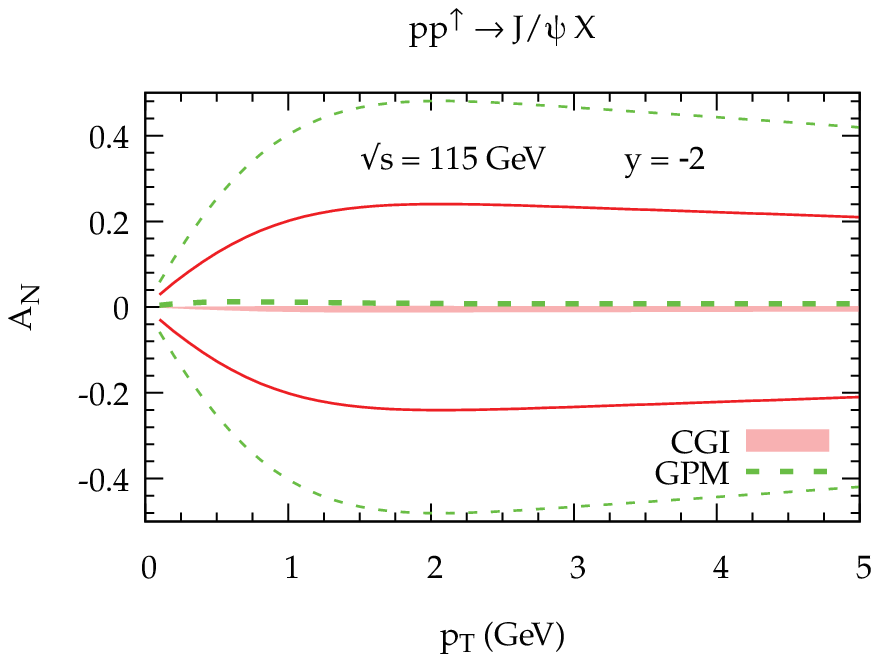}
\vspace*{-18truept}
\caption{Maximized and constrained $J/\psi$ Sivers SSAs for the GPM and CGI-PGM, compared to RHIC data~\cite{Aidala:2018gmp} (upper panels) and predicted for LHC fixed-target energies (lower panels). See text for details. }
\label{fig:piD}
\end{center}
\end{figure}

In Fig.~\ref{fig:max} we show the range of values for the $J/\psi$ Sivers SSA~\footnote{These results on the $J/\psi$ Sivers SSA update and supersede the corresponding ones of Ref.~\cite{DAlesio:2017rzj}.} allowed by the positivity bound compared with  PHENIX data \cite{Adare:2010bd}, both vs.~$p_T$ at fixed $x_F$ and vs. $x_F$ at fixed $p_T$. The largest green bands refer to the GPM, while the pale-blue ones to the CGI-GPM. Due to ISIs, the asymmetry for the CGI-GPM scheme is halved in size and opposite in sign as compared to the GPM one.
The purple solid lines in the right panels of  Fig.~\ref{fig:max}
show the GPM maximized value of the SSA reduced by a factor 20, and the constraining power of present PHENIX data for the $f$-type GSF.
Stronger phenomenological constraints for both the $f$-type and the $d$-type GSFs can be obtained by a combined analysis of RHIC SSA data for neutral pion production at central rapidity, $D$ meson and $J/\psi$ production. We will not discuss this study here (see Ref.~\cite{DAlesio:2018rnv} for details), limiting ourself to present, in Fig.~\ref{fig:piD}, some results for the $J/\psi$ Sivers SSA at RHIC~\cite{Aidala:2018gmp} and related predictions for future polarized fixed-target experiments at the LHC, like LHCSpin~\footnote{These results on the $J/\psi$ Sivers SSA update and supersede the corresponding ones of Ref.~\cite{DAlesio:2018rnv}.}.
In the upper panels of Fig.~\ref{fig:piD} we show, for both the GPM and the CGI-GPM cases, the maximized values of the $J/\psi$ Sivers asymmetry (left) and its value obtained constraining the GSFs by means of pion and $D$ meson production data (right), as a function of $x_F$ at fixed $p_T=1.65$ GeV and $\sqrt{s}=200$ GeV.
Notice the difference in the vertical scales between the two plots.
In the lower panels we show predictions for the maximized and the constrained $J/\psi$ Sivers asymmetry at a typical c.m.~energy for a fixed target experiment at LHC, $\sqrt{s}=115$ GeV, for both the GPM and CGI-GPM cases, as a function of $x_F$ at $p_T=2$ GeV (left) and as a function of $p_T$ at $y=-2$ (right).
To conclude, we summarize the current status of our ongoing project on the study of spin and azimuthal asymmetries for quarkonium production in $pp$ collisions within the TMD and NRQCD approaches.
We have presently completed the calculation of direct quarkonium production in (un)polarized $pp$ collisions in the low $p_T$ region.
A detailed phenomenological study is in progress. Some open points deserve further study: The dependence on color octet LDMEs; the regulation of divergences  in the very low $p_T$ range (TMD contributions can effectively  account for these effects, at least partially); the dependence on soft regulators for the hard cross sections. All these issues are less relevant for spin and azimuthal asymmetries, which are ratios of cross sections. This analysis and its extension to the CGI-GPM case are in progress and will be presented elsewhere.

\textit{Work partially supported by  Fondazione Sardegna under the project ``Quarkonium at LHC energies", CUP F71I17000160002 (University of Cagliari).
}

\vspace*{-2truept}
\providecommand{\href}[2]{#2}\begingroup\raggedright\endgroup


%

\end{document}